\documentclass[conference, a4paper,9pt]{IEEEtran}
\IEEEoverridecommandlockouts

\usepackage{amssymb}
\usepackage{cite, }
\usepackage{amsmath,amsfonts}
\usepackage{newtxmath}
\usepackage{graphicx}
\usepackage[dvipsnames]{xcolor}
\usepackage{textcomp}
\usepackage{threeparttable}
\usepackage{multirow}
\usepackage[linesnumbered,ruled,vlined]{algorithm2e}
\usepackage{svg}
\usepackage{caption}
\usepackage{subfigure}
\usepackage{hhline}
\usepackage{enumitem}
\usepackage{listings}
\usepackage{booktabs} 
\usepackage{tablefootnote}
\usepackage{afterpage}
\usepackage{colortbl}
\usepackage{cleveref}
\usepackage{soul}
\usepackage{amsmath, amssymb} 
\usepackage{pifont} 
\usepackage{afterpage}
\usepackage{float}
\usepackage{microtype}
\usepackage{cleveref}
\crefname{section}{§}{§§}
\Crefname{section}{§}{§§}
\usepackage{graphicx}
\usepackage{makecell}

\definecolor{mygreen}{RGB}{34,139,34}
\definecolor{myred}{RGB}{178,34,34}

\newcommand{\cmark}{\textcolor{mygreen}{\ding{51}}}
\newcommand{\xmark}{\textcolor{myred}{\ding{55}}}

\begin{document}
\pagestyle{empty}

\newcommand{\streamername}{\mbox{\textit{DataMaestro}}\xspace}
\newcommand{\streamersname}{\mbox{\textit{DataMaestro}s}\xspace}

\newcommand{\datastreamingunitname}{\mbox{DSE}\xspace}
\newcommand{\datastreamingunitsname}{\mbox{DSEs}\xspace}

\title{\streamername: A Versatile and Efficient Data Streaming Engine Bringing Decoupled Memory Access To Dataflow Accelerators }

\author{\normalsize{Xiaoling Yi$^{1*}$, Yunhao Deng$^{1*}$, Ryan Antonio$^{1}$, Fanchen Kong$^{1}$, Guilherme Paim$^2$}, Marian Verhelst$^{1}$ \\
	\normalsize{$^1$MICAS-ESAT, KU Leuven, Leuven, Belgium}\\
	\normalsize{$^2$INESC-ID, Instituto Superior Técnico, Universidade de Lisboa, Lisboa, Portugal}\\
	\vspace{-0.9cm}
	\thanks{*Equal contribution.}
}

\maketitle

\begin{abstract}

Deep Neural Networks (DNNs) have achieved remarkable success across various intelligent tasks but encounter performance and energy challenges in inference execution due to data movement bottlenecks. We introduce \streamername, a versatile and efficient data streaming unit that brings the decoupled access/execute architecture to DNN dataflow accelerators to address this issue.
\streamername supports flexible and programmable access patterns to accommodate diverse workload types and dataflows, incorporates fine-grained prefetch and addressing mode switching to mitigate bank conflicts, and enables customizable on-the-fly data manipulation to reduce memory footprints and access counts. We integrate five \streamersname with a Tensor Core-like GeMM accelerator and a Quantization accelerator into a RISC-V host system for evaluation. The FPGA prototype and VLSI synthesis results demonstrate that \streamername helps the GeMM core achieve nearly $100\%$ utilization, which is $1.05$-$21.39\times$ better than state-of-the-art solutions, while minimizing area and energy consumption to merely $6.43\%$ and $15.06\%$ of the total system. 

\end{abstract}

\section{Introduction}
\label{sec:intro}

\begin{table*}[t]
	\centering
	\begin{threeparttable}
		\caption{Comparison of the SotA data movement solutions with \streamername.}
		\label{tab: sota_feature_compare}
		\setlength{\tabcolsep}{3pt} 
		\renewcommand{\arraystretch}{1.1} 
		\small
		\begin{tabular}{c|c|c|c|c|c|c|c|c|c}
			\toprule
			                                  & Gemmini \cite{genc2021gemmini} & BitWave \cite{shi2024bitwave} & \cite{schneider2024energy} & FEATHER \cite{tong2024feather} & SSR \cite{schuiki2020stream} & HWPE \cite{hwpe2025} & Buffet \cite{pellauer2019buffets} & Softbrain \cite{nowatzki2017stream} & \streamername    \\
			\midrule
			Open source                       & \cmark                         & \xmark                        & \xmark                     & \cmark                         & \cmark                       & \cmark               & \cmark                            & \xmark                              & \cmark           \\
			Reusable design                   & \xmark                         & \xmark                        & \xmark                     & \xmark                         & \xmark                       & \cmark               & \cmark                            & \xmark                              & \cmark           \\
			Decoupled access/execute          & \xmark                         & \xmark                        & \xmark                     & \xmark                         & \cmark                       & \cmark               & \cmark                            & \cmark                              & \cmark           \\
			Programmable affine access        & \cmark  ($2$-$D$)              & \xmark                        & \cmark ($2$-$D$)           & \xmark                         & \cmark ($4$-$D$)             & \cmark ($3$-$D$)     & \cmark ($2$-$D$)                  & \cmark ($2$-$D$)                    & \cmark ($N$-$D$) \\
			Fine-grained prefetch             & \xmark                         & \xmark                        & \xmark                     & \xmark                         & \xmark                       & \cmark               & \xmark                            & \xmark                              & \cmark           \\
            \makecell[c]{Runtime addressing \\ mode switching} & \xmark & \xmark & \xmark & \xmark & \xmark & \xmark & \xmark & \xmark & \cmark \\
			On-the-fly data manipulation      & \xmark                         & \xmark                        & \xmark                     & \cmark                         & \xmark                       & \xmark               & \xmark                            & \xmark                              & \cmark           \\
			\bottomrule
		\end{tabular}
	\end{threeparttable}
	\vspace{-0.3cm}
\end{table*}

Deep Neural Networks (DNNs) have significantly gained popularity for tackling complex intelligence tasks and potentially advancing humanity toward the Artificial General Intelligence (AGI) era.
To meet the stringent performance and energy consumption requirements of DNN deployment, numerous contemporary domain-specific dataflow accelerators have been developed~\cite{shi2024bitwave, pouya2023diana, norrie2020google, du202328nm, nowatzki2017stream, talpes2020compute}, incorporating extensive optimizations in the processing element (PE) array design and dataflow strategies.
However, as DNN workloads and PE arrays continue to scale, data communication between the accelerator datapath and the memory hierarchy emerges as a critical bottleneck in DNN dataflow accelerators \cite{tong2024feather, kwon2018maeri, huang2024mind, parashar2019timeloop, mei2021zigzag, hameed2010understanding}.

Despite various solutions having been proposed to address this bottleneck, three shortcomings hinder their widespread deployment toward efficient DNN dataflow accelerators:
\textbf{1).~Non-Reusable Design.}
Data movement units are typically implemented as dedicated hardware blocks, tightly coupled to a specific accelerator, featuring fixed access patterns and bandwidths tuned to specific workloads \cite{shi2024bitwave, pellauer2019buffets, nowatzki2017stream, tong2024feather, schneider2024energy}. Given the diverse sizes, topologies, and rapid evolution of DNN workloads, these dedicated data movement units lack the versatility to adapt to different dataflows' requirements, leading to suboptimal overall system performance and energy efficiency. For instance, BitWave~\cite{shi2024bitwave} achieves very high PE array utilization for convolution operations in convolutional neural networks (CNNs) by leveraging specialized optimizations but falls short in general matrix-matrix multiplication (GeMM), which is pervasive in Transformer models. Moreover, the inability to reuse these dedicated data movement blocks across different accelerators results in redundant engineering efforts and hinders competitive time-to-market, especially considering the rapid evolution of DNNs and next-generation computing platforms.
\textbf{2).~Performance Decrease Caused by Bank Conflicts.}
The on-chip memory bank conflict challenge is a primary obstacle to achieving peak system performance.
For instance, Gemmini~\cite{genc2021gemmini} exhibits a PE array utilization as low as $10\%$ due to the lack of bank conflict management in the access of the different operands.
To address this problem, many accelerators \cite{pouya2023diana, tortorella2023redmule, shi2024bitwave, liao2021ascend} leverage private on-chip memory for each operand, such as dedicated input, weight, and output buffers. However, bank conflicts still arise when data needed in a single cycle is stored in different wordlines within the same bank, a scenario commonly encountered in strided convolution operations. Moreover, it imposes additional limitations on the tiling of the workload to meet the requirement of the smallest memory and introduces extra complexity to manage multiple address spaces \cite{mei2021zigzag}.
\textbf{3).~Expensive Data Manipulation.}
Many accelerators require preprocessing of operand before computation, including permutation, data alignment, and \textit{im2col}.
A common approach is to design standalone data manipulation units to perform these data rearrangement operations \cite{norrie2020google, talpes2020compute, genc2021gemmini}. These units require additional memory accesses and intermediate data storage, which again increases the likelihood of memory conflicts and elevates memory energy consumption.
In conclusion, there is an absence of a versatile and efficient data movement solution to address the diverse dataflow and data manipulation requirements of different accelerators while also ensuring high system performance.

The decoupled access/execute (DAE) architecture \cite{smith1982decoupled} separates data access and computation processes into two independent streams to prevent mutual stalls. This approach has been applied in numerous works \cite{domingos2021unlimited, schuiki2020stream, wang2019stream}, demonstrating performance speed-ups of $2$-$5\times$ for linear algebra workloads \cite{schuiki2020stream, domingos2021unlimited}.
The DAE architecture is well-suited for DNN workloads, as its regular multidimensional access patterns \cite{sze2017efficient} enable data to be fetched ahead of computation. This facilitates continuous data stream consumption and production from the perspective of PE arrays, helping the accelerator system reach its theoretical peak performance.
Previous works \cite{nowatzki2017stream, pellauer2019buffets} have successfully applied the DAE architecture in accelerator designs. However, they fail to meet the diverse dataflow, data access patterns, and high bandwidth requirements of DNN workloads and the aforementioned challenges, rendering them suboptimal for DNN dataflow accelerators.

In this paper, we present a versatile and efficient data streaming engine, called \streamername \footnote{\streamername and the \streamername-boosted accelerator system are open-sourced at: https://github.com/KULeuven-MICAS/snax\_cluster.}, which brings decoupled memory access to DNN dataflow accelerators. \streamername is highly optimized and can be reused across different dataflow accelerators with flexible and programmable access patterns while achieving nearly $100\%$ PE array utilization. A comprehensive comparison between \streamername and state-of-the-art (SotA) solutions is listed in Table \ref{tab: sota_feature_compare}.
In summary, our main contributions lie in:

\begin{itemize}
	\item We present \streamername, a generalized, versatile, and efficient data streaming engine for diverse DNN dataflow accelerators, with design-time and runtime configurability, enabling decoupled memory access for arbitrary dataflows (\cref{sec:arch_overview} and \cref{sec:agu}).
	\item We introduce fine-grained prefetch and runtime-configurable addressing mode switching features inside \streamername, which minimizes bank conflicts and fully exploits the available on-chip memory bandwidth (\cref{sec:memory_interface} and \cref{sec:address_remap}).
	\item We provide on-the-fly data manipulation capability inside the \streamername by customizable datapath extensions to effectively reduce memory footprints and access counts (\cref{sec:extension}).
	\item We demonstrate the versatility and efficiency of \streamername by integrating it with a Tensor core \cite{nvidia2018turing}-like GeMM accelerator and a Quantization accelerator into a RISC-V host system. The FPGA prototype and VLSI synthesis results show that \streamername enhances the GeMM PE array utilization to $95.45\%$-$99.98\%$ under real-world DNN workloads, boosting normalized throughput to $1.05$-$21.39 \times$ over SotA, while only consuming $6.43\%$ and $15.06\%$ of system area and energy (\cref{sec: eval}).
\end{itemize}

\section{Background}

\label{sec:bg}
\subsection{Dataflow and Data Layout}
\label{sec:bg_df}

Typical DNN kernels, such as GeMM, multi-head attention, convolution, and pooling, can be represented as nested loops that iterate over the input, weight, and output tensor dimensions. Dataflow optimizations, like splitting and reordering these loops, open up a vast mapping space for efficient hardware processing \cite{parashar2019timeloop, mei2021zigzag}. Specifically, these optimizations include spatial unrolling (SU) for parallel execution and temporal unrolling (TU) to control data tiling and the order of loop iterations. 

The data access pattern, which specifies the data required by the accelerator in each clock cycle, is inherently determined by the dataflow. 
However, the data layout, referring to the organization of tensor data in memory, also influences data access patterns.
A proper data layout is crucial for optimizing DNN performance, as it directly impacts memory access efficiency and can lead to PE array underutilization if mismatched with the dataflow \cite{ye2020hybriddnn, tong2024feather}.

\subsection{Decoupled Access/Execute Architecture for Dataflow Accelerators}
\label{sec:bg_dae}

\begin{figure}[t]
    \centering
    \includegraphics[width=0.95\linewidth]{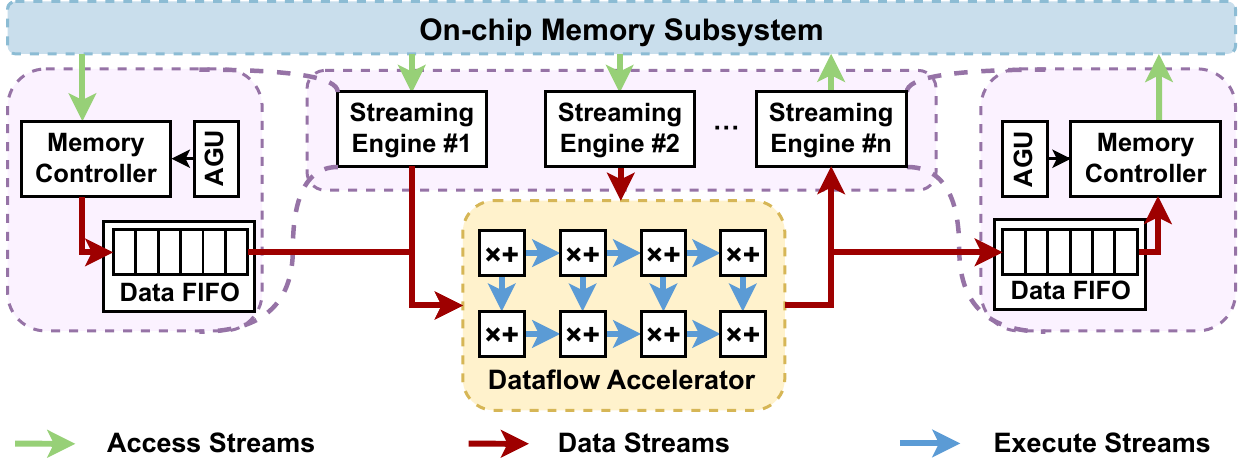}
    \caption{Overview of decoupled access/execute architecture.}
    \label{fig:dae}
    \vspace{-0.5cm}
\end{figure}

A decoupled access/execute (DAE) architecture \cite{smith1982decoupled, nowatzki2017stream, pellauer2019buffets} explicitly decouples memory access and computation by introducing data streaming engines between them, as illustrated in Figure \ref{fig:dae}.
This architecture integrates three key components — memory subsystem, data streaming engine, and accelerator datapath—which collaborate coherently to produce/consume three types of streams: access streams, data streams, and execute streams. The memory subsystem serves as a data reservoir, responding to the memory requests via the access stream. The accelerator's datapath applies computation to the data streams and produces results via the execute stream.
The data streaming engine bridges the memory subsystem with the accelerator datapath, orchestrating data flow between them. It transforms scattered data in memory which is arranged in a specific format (data layout), into a continuous data stream, which is organized in the format required by the accelerator (data flow) and vice versa, by interacting with the memory through the access stream.

Typically, the streaming engine consists of three main components: an Address Generation Unit (AGU), a memory interface controller, and a data FIFO.
The AGU generates address sequences, which are then consumed by the memory interface controller to perform memory operations. The data FIFO temporarily stores data streams, decoupling memory access from computation. This decoupling enables data prefetch operations and thus minimizes computation interruptions caused by memory stalls. Such an approach is particularly beneficial for DNN workloads due to their regular and deterministic data access patterns. By incorporating this architecture, a harmonious access-data-execute stream orchestration is established within the accelerator system. 
In the next section, we present \streamername, a highly optimized data streaming engine designed for DNN accelerators, which efficiently streamlines the data streams to enhance the performance of dataflow accelerators. 

\section{\streamername Architecture}

\begin{figure*}[t]
    \centering
    \includegraphics[width=0.9\textwidth]{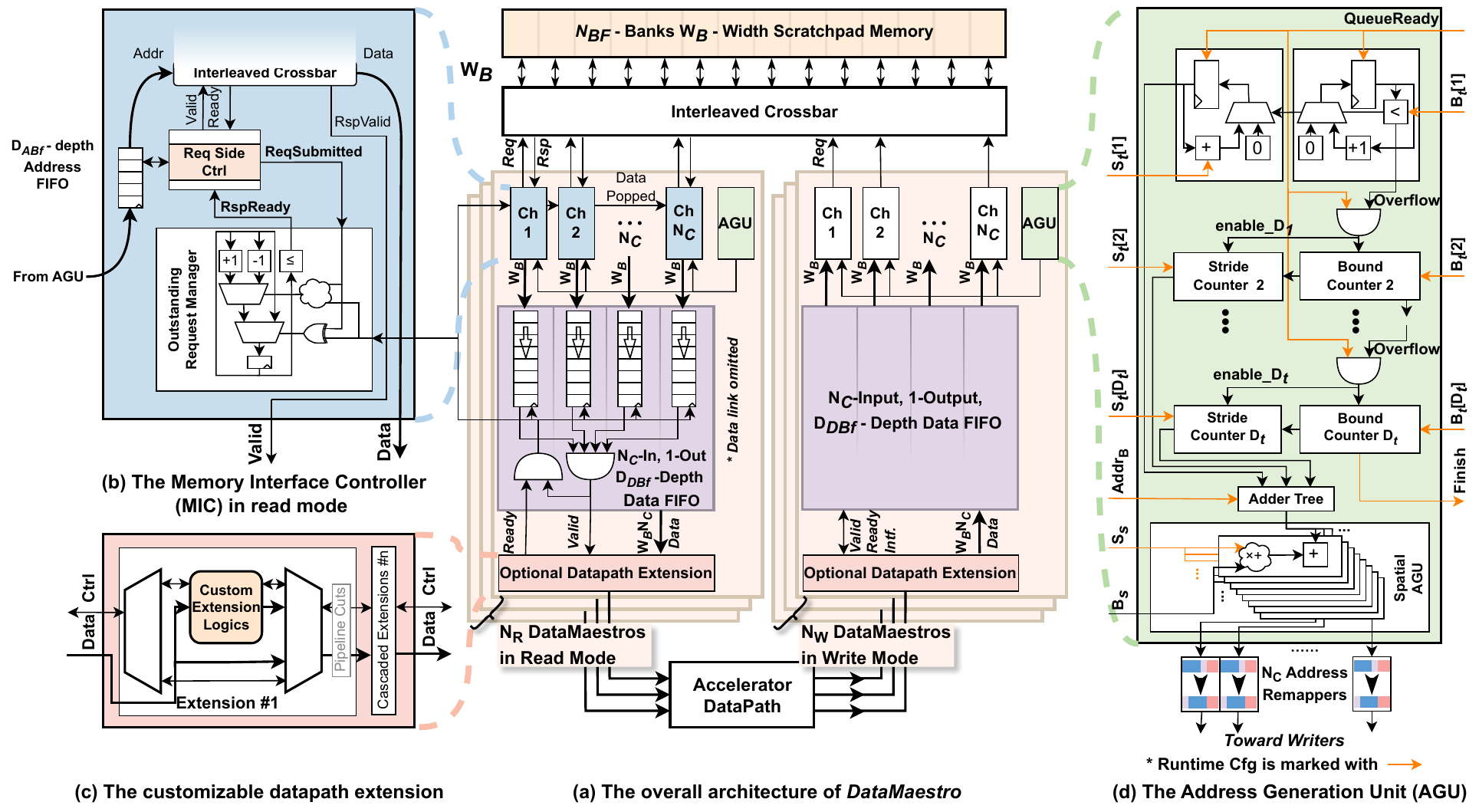}
    \caption{\streamername architecture.}
    \label{fig:streamer-main-graph}
    \vspace{-0.5cm}
\end{figure*}

In this section, we will first provide an overview of the \streamername architecture in \cref{sec:arch_overview}, along with all of its configurable parameters, which are listed in Table \ref{tab:cfg}. Next, we introduce the three novel architectural features that maximize \streamername's efficiency and minimize memory bank conflicts across a wide range of workloads: 1) highly configurable and area-efficient N-Dimensional ($N$-$D$) address generation unit (\cref{sec:agu}); 2) fine-grained prefetch (\cref{sec:memory_interface}); and 3) runtime-configurable addressing mode switching (\cref{sec:address_remap}). Finally, we discuss the datapath extension interface of \streamername, which enables customized on-the-fly data manipulation (\cref{sec:extension}). These unique features of \streamername are key to its high efficiency and enhancement of the performance of DNN dataflow accelerators.

\begin{table}[ht]
    \centering
    \begin{threeparttable}
        \caption{Design-time parameters and runtime configurations in \streamername.}
        \label{tab:cfg}
        \begin{tabular}{c|c|c|c}
            \arrayrulecolor{black}
            \toprule
            Scope & Parameter & Description & Type \tnote{\textdagger} \\
            \midrule
            \multirow{12}{*}{Design Time} 
            & $N_{R}$ & Num. of read \streamername & Int \\
            & $N_{W}$ & Num. of write \streamername & Int \\
            \cline{2-4}
            & $Mode_{R/W}$ & Read or Write mode & Bool. \\
            & $B_{s}$ & Spatial Bounds & [Int] \\
            & $D_{s}$ & Num. of Spatial Dimensions & Int \\
            & $D_{t}$ & Num. of Temporal Dimensions & Int \\
            & $N_C$ & Num. of Channels & Int \\
            & $D_{ABf}$ & Address Buffer Depth & Int \\
            & $D_{DBf}$ & Data Buffer Depth & Int \\
            & $DP_{ext}$ & Datapath Extensions & [Ext.] \\
            & $W_{B}$ & Bank Width & Int \\
            & $N_{BF}$ & Num. of Bank in Full Memory & Int \\
            & $N_{BG}$ & Num. of Bank in One Group & [Int] \\
            \midrule
            \multirow{6}{*}{Run Time}
            & $Addr_{B}$ & Base Address & UInt \\
            & $S_{s}$ & Spatial Strides & [UInt] \\
            & $B_{t}$ & Temporal Bounds & [UInt] \\
            & $S_{t}$ & Temporal Strides & [UInt] \\
            & $R_S$ & Addressing Mode Selection & UInt \\
            \bottomrule
        \end{tabular}
        \begin{tablenotes}
            \footnotesize
            \item \textdagger \:Type enclosed in square brackets {[\;]} means List.
        \end{tablenotes}
        \renewcommand{\arraystretch}{1.0}
    \end{threeparttable}
    \vspace{-0.5cm}
\end{table}

\subsection{Architecture Overview of \streamername}

\label{sec:arch_overview}
Figure \ref{fig:streamer-main-graph} (a) illustrates a scenario where $N_{R}$+$N_{W}$ \streamersname are connected to a multi-banked scratchpad memory subsystem (top) and an accelerator (bottom), effectively managing data streams between them to create a complete accelerator system. One read \streamername (middle, left) and one write \streamername (middle, right) are depicted in detail.
The memory subsystem consists of an $N_{B}$-banked scratchpad memory that provides high memory bandwidth, along with an interleaved crossbar that ensures full accessibility from any request port.
To meet the specific requirements of different accelerator ports, $N_{R}$ read and $N_{W}$ write \streamersname are deployed, each of which can be configured independently at both design time and runtime, by providing the parameters listed in Table \ref{tab:cfg}.

A read or write \streamername primarily includes $N_{C}$ independent memory interaction channels, breaking one wide memory request into multiple narrower operations for asynchronous, fine-grained memory access. Each channel is equipped with a dedicated Memory Interface Controller (MIC) (Figure \ref{fig:streamer-main-graph} (b)) and a data FIFO. The MIC consumes addresses from the address generation unit (AGU) (Figure \ref{fig:streamer-main-graph} (d)) to issue memory requests, and the data FIFO acts as a buffer between the \streamername and the accelerator. This facilitates a decoupled access/execute architecture for dataflow accelerators to effectively hide memory latency. At the output ports of the FIFOs, data fetched by multiple channels are gathered into a wide data stream and sent as a single wide data word to the accelerator datapath, enabling highly paralleled execution. 
Furthermore, at the interface between the \streamername and the accelerator (Figure \ref{fig:streamer-main-graph} (c)), multiple user-customizable datapath extensions can optionally be inserted to perform on-the-fly data processing in a cascaded manner without requiring intermediate memory storage. In the following subsections, we will delve into each part of \streamername in more detail.

\subsection{Programmable Affine Access Pattern by Address Generation Unit}
\label{sec:agu}

\begin{figure}[t]
    \centering
    \includegraphics[width=0.95\linewidth]{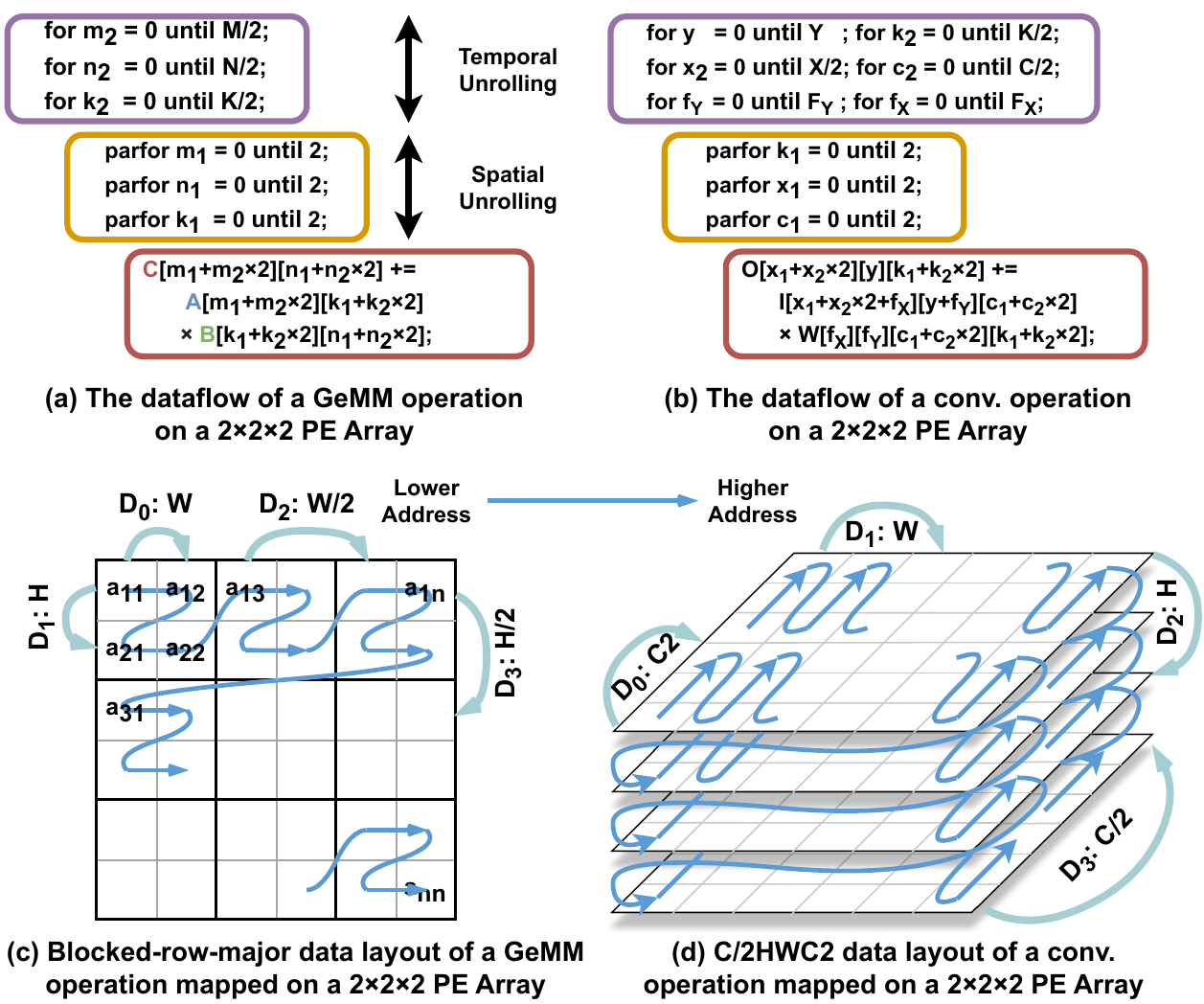}
    \caption{Dataflow and data layout of GeMM and convolution operations mapped on $2\times2\times2$ PE array.}
    \label{fig:gemm_conv_df}
    \vspace{-0.2cm}
\end{figure}

The data access pattern for DNN workloads, determined by the dataflow and data layout, can typically be represented as an affine pattern \cite{sze2017efficient}.
Furthermore, DNN workloads encompass a wide range of operation types (e.g., GeMM and various types of convolutional operations) and tensor shapes (e.g., height, width, and channel dimensions) \cite{shi2023cmds}, necessitating data access patterns with high flexibility. For example, as illustrated in Figure \ref{fig:gemm_conv_df}, the dataflow and data layout of GeMM and convolution workloads mapped on a simple $2\times2\times2$ PE array already differ significantly. For GeMM workload, the matrix $A$ is stored using a $4$-$D$ block-row-major data layout \cite{lam1991cache, zheng2020optimizing} (Figure \ref{fig:gemm_conv_df}~(c)), and exhibits a $3$-$D$ access pattern across different cycles (Figure \ref{fig:gemm_conv_df}~(a)). In contrast, for convolution, the input tensor is stored using a $4$-$D$ blocked data layout (Figure \ref{fig:gemm_conv_df}~(d)), i.e., $C/2HWC2$, and features a $6$-$D$ data access pattern (Figure \ref{fig:gemm_conv_df}~(b)). The diverse access pattern (while remaining affine) in DNN workloads motivates the design of a flexible and programmable AGU within \streamername.

\begin{figure}[t]
    \centering
    \includegraphics[width=0.95\linewidth]{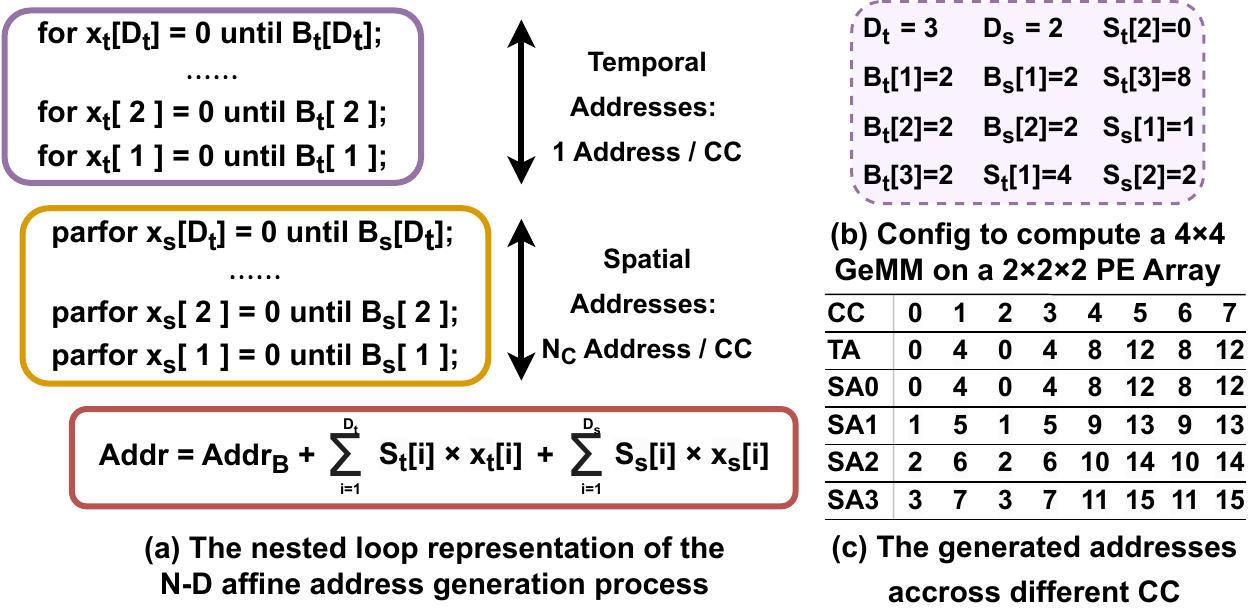}
    \caption{The nested loop representation of the $N$-$D$ affine address generation in \streamername's AGU and a simple address generation example for a $M$=$N$=$K$=$4$ GeMM workload mapped on a $2\times2\times2$ PE array.}
    \label{fig:agu_for_loop}
    \vspace{-0.3cm}
\end{figure}

The AGU in \streamername enables $N$-Dimensional ($N$-$D$) programmable affine address generation as shown in Figure \ref{fig:agu_for_loop} (a), using symbols from Table \ref{tab:cfg}. This affine function effectively describes the mapping of the $N$-$D$ data access space to the $1$-$D$ address space. Our address generation process involves sequentially generating temporal addresses (TA) to represent temporal data access patterns and simultaneously generating multiple spatial addresses (SA) based on every temporal address for parallel data access, satisfying the requirements of temporal and spatial unrolling for a specific dataflow. A detailed example of the AGU configuration for an $M$=$N$=$K$=$4$ GeMM workload mapped on a $2\times2\times2$ PE array is presented in Figure \ref{fig:agu_for_loop} (b), with the corresponding address generation process depicted in Figure \ref{fig:agu_for_loop} (c). 

Directly using a single counter to track the address generation progress necessitates modulo units and dividers to compute loop indices, followed by multipliers and adders to calculate temporal addresses. However, this leads to long combinatorial paths and significant hardware overheads that linearly scale with the number of dimensions. In \streamername, we incorporate microarchitectural optimizations to enable high-dimensional address generation with low overhead and shorter combinatorial paths. As shown at the top of Figure \ref{fig:streamer-main-graph} (d), each dimension of the temporal AGU is implemented in a dual-counter structure, including a simple counter (Bound Counter) to record the loop index and a step-programmable counter (Stride Counter) to generate the temporal address offset. 
Then, address offsets of all dimensions are summed with the base address ($Addr_{B}$) and consumed by a multi-channel spatial AGU (the bottom of Figure \ref{fig:streamer-main-graph} (d)), which generates distinct addresses for each channel. Based on this optimized hardware architecture, \streamername supports specifying arbitrary temporal address loop dimension, spatial address loop dimension, and spatial address loop bounds at design time, while the strides for both temporal and spatial addresses can be programmed at runtime, providing full flexibility to accommodate diverse data access patterns in DNN dataflow accelerators.

\subsection{Fine-grained Prefetch by Memory Interface Controller}
\label{sec:memory_interface}

\streamername conducts fine-grained prefetch by breaking one wide memory request into multiple narrower channels and issuing them independently. During this process, Memory Interface Controllers (MICs) in read mode play a key role in sending fine-grained requests, gathering response data, and storing it in the data FIFO. 
The MIC in read mode includes an Outstanding Request Manager (ORM) and a Request Side Controller (RSC), as shown in Figure \ref{fig:streamer-main-graph}(b). The ORM tracks the utilization of the data FIFO and reserves the FIFO slots for in-flight memory requests, throttling the RSC if no available slots in the FIFO can be reserved. With permission from the ORM and valid addresses from the AGU, the RSC immediately issues new memory requests. This aggressive and fine-grained data prefetch mechanism maximizes memory bandwidth utilization.

\vspace{-0.1cm}
\subsection{Low Overhead Addressing Mode Switching by Address Remapper}
\label{sec:address_remap}

\begin{figure}
    \centering
    \includegraphics[width=0.9\linewidth]{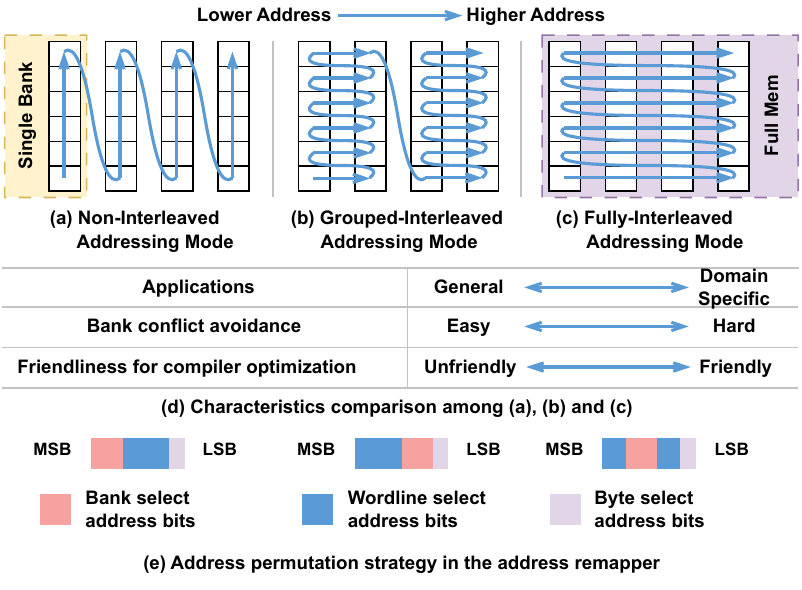}
    \caption{Addressing modes supported by \streamername.}
    \label{fig:addressing}
    \vspace{-0.5cm}
\end{figure}

Two addressing modes are commonly used in computer systems for accessing multi-banked memory: Fully-Interleaved Memory Addressing (FIMA), where addresses are interleaved across all banks (Figure \ref{fig:addressing}(a)), and Non-Interleaved Memory Addressing (NIMA), where contiguous addresses are allocated to a single bank (Figure \ref{fig:addressing}(c)). Between these two lies an intermediate mode, termed Grouped-Interleaved Memory Addressing (GIMA) by us, where some banks are grouped, and the address is interleaved intra-group and remains contiguous inter-group (Figure \ref{fig:addressing}(b)). The characteristics of these three addressing modes are summarized in  \ref{fig:addressing} (d). 
While FIMA is widely used in general-purpose computing systems \cite{zhang2000permutation}, it is more prone to bank conflicts in domain-specific accelerator (DSA) systems. As a result, contemporary dataflow accelerators often favor the NIMA mode \cite{genc2021gemmini, shi2024bitwave, schneider2024energy}. 
However, the compiler needs to carefully allocate data for maximal performance and it constrains the tilings of the workload to meet the smallest memory requirement.
Positioned between these two, GIMA strikes a trade-off between them. To accommodate diverse DNN workloads and dataflow accelerators, \streamername is designed to support all three addressing modes, with the capability to switch between them at runtime, offering the designer an extra degree of freedom to optimize the data allocation.

The main challenge of addressing mode switching is the overhead introduced in address decoding. However, addressing mode switching can be represented as a simple bit permutation of the address when the bank number in one group is a power of two. Based on this insight, \streamername leverages a memory address remapper that facilitates simple address bit permutation for addressing mode switching, as illustrated in Figure \ref{fig:addressing} (e). 
\streamername instantiates the bit permutations in hardware based on the number of banks in total ($N_{BF}$) and the number of banks in one group ($N_{BG}$) at design time. Then, the permuted addresses are connected to a multiplexer, allowing the user to select the addressing mode by setting $R_S$ at the run time.

\subsection{On-the-fly Data Manipulation by Datapath Extension}
\label{sec:extension}
\streamername provides a mechanism to insert customizable extensions, such as data quantization and permutation, between \streamername and accelerator datapath in a plug-and-play manner to conduct on-the-fly data manipulation, as shown in Figure \ref{fig:streamer-main-graph} (c). 
Datapath extensions are instantiated based on the design-time configuration $DP_{ext}$, with the output of one extension serving as the input for the next. 
Furthermore, the logic to bypass any extension is automatically inserted so that users can disable any extension at runtime. 

\section{evaluation}
\label{sec: eval}
\subsection{Setup and Methodology}
\label{sec:exp_setup}
\begin{figure}
    \centering
    \begin{minipage}{0.48\linewidth}
        \centering
        \includegraphics[width=\linewidth]{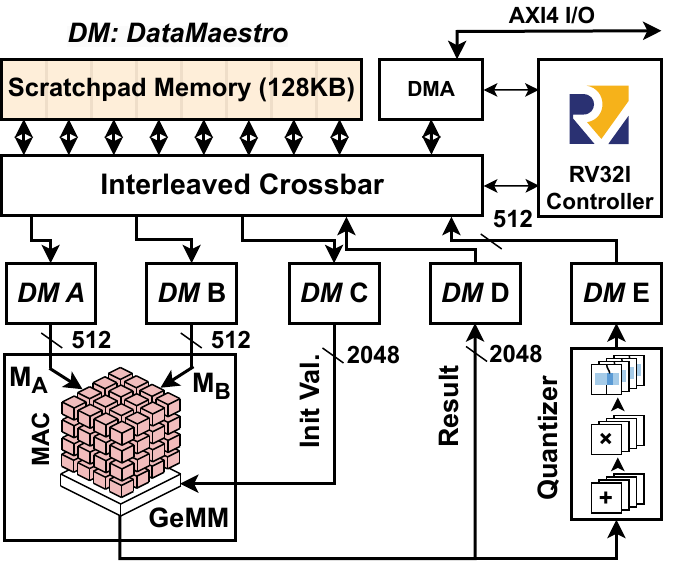}
    \end{minipage}
    \begin{minipage}{0.5\linewidth}
        \scalebox{0.6}{
            \large{
                \begin{threeparttable}
                    \centering
                    \begin{tabular}{c|c|c|c|c|c}
                        \toprule
                        Parameter  & $A$                             & $B$      & $C$   & $D$   & $E$ \\
                        \midrule
                        \arrayrulecolor[gray]{0.8}
                        $D_{spat}$ & 1                               & 1        & 2     & 2     & 1   \\
                        \hline
                        $B_{spat}$ & [8]                             & [8]      & [8,4] & [8,4] & [8] \\
                        \hline
                        $D_{temp}$ & 6                               & 3        & 3     & 3     & 3   \\
                        \hline
                        $W_{B}$    & 64                              & 64       & 64    & 64    & 64  \\
                        \hline
                        $N_{C}$    & 8                               & 8        & 32    & 32    & 8   \\
                        \hline
                        $D_{DBf}$  & 8                               & 8        & 1     & 1     & 1   \\
                        \hline
                        \arrayrulecolor{black}
                        $DP_{ext}$ & \multicolumn{2}{c|}{[Trans.]}   & [Broad.] & [\;]  & [\;]        \\
                        \arrayrulecolor[gray]{0.8}
                        \hline
                        $N_{BF}$   & \multicolumn{5}{c}{2048}                                         \\
                        \hline
                        $N_{BG}$   & \multicolumn{5}{c}{[2048, 512]}                                  \\
                        \arrayrulecolor{black}
                        \bottomrule
                    \end{tabular}
                \end{threeparttable}
            }
        }
    \end{minipage}
    \caption{The \streamername evaluation system (left) and \streamersname design-time parameters used in it (right).}
    \label{fig:case_study_arch}
    \vspace{-0.5cm}
\end{figure}

To assess the versatility and efficiency of \streamername, we integrate it with a Tensor Core \cite{nvidia2018turing}-like GeMM accelerator and a Quantization accelerator into a RISC-V host system~\cite{zaruba2020snitch}, as depicted in Figure \ref{fig:case_study_arch} (left). The GeMM accelerator features a $3$-$D$ 8$\times$8$\times$8 PE array, capable of executing workloads expressed as $D_{32} = A_{8} \otimes B_{8} + C_{32}$, where the subscripts denote operand data precision and $\otimes$ signifies either GeMM or convolution operations, depending on the runtime configuration of \streamersname. The output from the GeMM accelerator can be directed either to the Quantization accelerator for post-processing, represented as $E_{8} = Rescale(D_{32})$, or sent back to the memory.
The GeMM and Quantization accelerators feature five data stream ports in total, each with different data access patterns and bandwidth requirements, served by five \streamername, as shown in Figure \ref{fig:case_study_arch} (right). The $6$-$D$ temporal AGU within \streamername $A$ facilitates implicit \textit{im2col} transformation \cite{zhou2021characterizing} for convolution operations. Two datapath extensions are implemented: 1). Transposer, which performs on-the-fly matrix tile transposition, and 2). Broadcaster, which duplicates data across channels, particularly useful for DNN workloads with per-channel quantization. Additionally, a customized compiler is developed to generate runtime configurations for these \streamersname, considering workload specifications and tensor data layouts.

We implement \streamername, GeMM, and Quantization accelerators at the register-transfer level (RTL) using Chisel \cite{bachrach2012chisel}. An ablation study is conducted using Verilator for cycle-accurate RTL simulation to evaluate utilization gains and memory access count reductions provided by each feature of \streamername (\cref{sec: util_eval}). Furthermore, an FPGA prototype is deployed on the AMD Versal\textsuperscript{\texttrademark} VPK180 FPGA to benchmark model-wise performance on real-world DNN workloads (\cref{sec: fpga_proto}).
The evaluation system is also synthesized using the Synopsys Design Compiler\textsuperscript{\textregistered} with GlobalFoundries 22FDX\textsuperscript{\textregistered} technology, operating at 1GHz and 0.8V with strict timing closure. Post-synthesis simulation is performed for power consumption analysis using Siemens QuestaSim\textsuperscript{\texttrademark} and Synopsys PrimeTime\textsuperscript{\textregistered} (\cref{sec: power_area_break}).

\subsection{Ablation Study for \streamername Evaluation}

\label{sec: util_eval}
\subsubsection{Workload and Evaluation Architecture Setting}
We conduct an ablation study to evaluate the effectiveness of each innovation in \streamername using $260$ different synthetic DNN workloads, categorized into three groups: 1) GeMM, 2) transposed GeMM, and 3) convolution. The synthetic benchmark set features various matrix sizes for GeMM and transposed GeMM, along with diverse feature map sizes, channels, kernel sizes, and strides for convolution, effectively representing typical Transformer and CNN layers.
Initially, we disable all features of \streamersname, architecturally similar to plain data movement units, inside the accelerator system as the baseline (\ding{172}).
Subsequently, we gradually introduce each feature, eventually forming an accelerator system with fully-featured \streamersname (\ding{173} to \ding{177}).

\subsubsection{Result Analysis}
\label{sec:result_analysis}
\begin{figure}
    \centering
    \includegraphics[width=\linewidth]{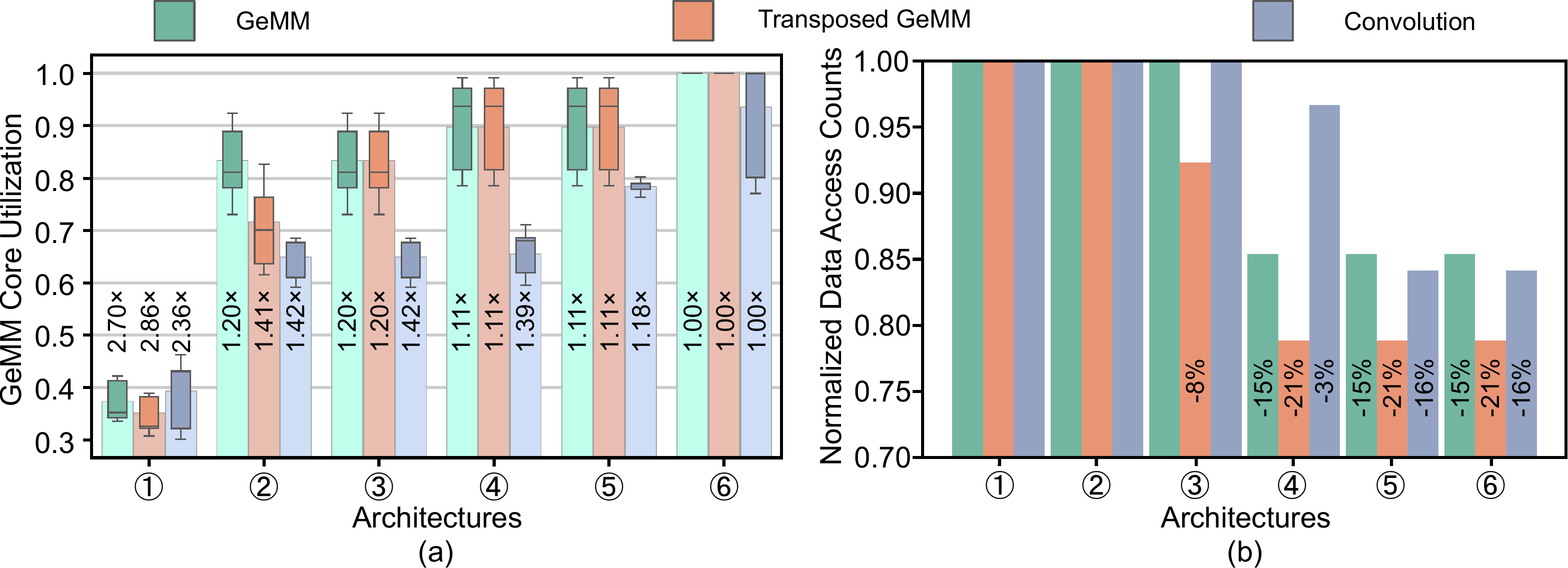}
    \caption{Ablation study results of GeMM core's utilization distribution presented in box plot, with average utilization annotated in bar chart (a), and normalized data access counts (b) across architectures with different \streamername features under the synthetic DNN workload set. \ding{172} = Baseline (all features turned off); \ding{173} = \ding{172} + Fine-grained prefetch; \ding{174} = \ding{173} + Transposer; \ding{175} = \ding{174} + Broadcaster; \ding{176} = \ding{175} + Implicit \textit{im2col}; \ding{177} = \ding{176} + Addressing mode switching.}
    \label{fig: synthetiic_utilization}
    \vspace{-0.5cm}
\end{figure}

Figure \ref{fig: synthetiic_utilization} (a) presents the GeMM core utilization distribution for each DNN kernel group using a box plot, with the average utilization inside each group depicted in the accompanying bar chart. As demonstrated in the results, \ding{173} achieves a significant increase in GeMM core utilization, ranging from $1.65 \times$ to $2.21 \times$ higher than the baseline \ding{172} across all workloads, thanks to fine-grained asynchronous prefetch.
Among all features, the Transposer is particularly effective for the transposed GeMM workload (\ding{174} compared to \ding{173}), improving utilization by $1.16 \times$, while Broadcaster increases utilization of all workloads by up to $1.09 \times$ (\ding{175} compared to \ding{174}).
The implicit \textit{im2col} \cite{zhou2021characterizing} transformation further delivers a $1.19 \times$ utilization improvement for convolution workloads (\ding{176} compared to \ding{175}). Finally, with addressing mode switching (\ding{177}), the utilization for the two GeMM workloads reaches $100\%$, with minimal variation across all matrix sizes.
For convolution workloads, the average utilization reaches $92.03\%$ but the utilization distribution exhibits notable variation across different workload sizes. Upon further analysis, this fluctuation is attributed to strided convolution layers, which require fetching noncontiguous input data across wider address ranges, resulting in unavoidable bank conflicts. Fortunately, strided convolution layers, typically used for feature map downsampling, comprise only a small portion of real-world DNN workloads and thus have a limited impact, as demonstrated in \cref{sec:dnn_benchmark}.
Meanwhile, datapath extensions for on-the-fly data manipulation significantly reduce redundant memory accesses, mitigating bank conflicts and saving energy. As illustrated in Figure \ref{fig: synthetiic_utilization} (b), Transposer reduces the data access counts by $15.86\%$ for transposed GeMM workloads (\ding{174} compared to \ding{173}), and Broadcaster effectively reduces memory accesses by up to $14.58\%$ across all three workloads (\ding{175} compared to \ding{174}).
Overall, the fully-featured \streamername (\ding{177}) achieves up to $2.89\times$ performance speedup and up to $21.15\%$ reduction in memory accesses across all test cases, compared to the baseline (\ding{172}). These results highlight that \streamername enables versatile and highly efficient data streaming to the accelerator across a variety of DNN kernels with different sizes.

\subsection{Real-world Neural Network Evaluation on FPGA}
\label{sec: fpga_proto}

\subsubsection{Evaluation Setup}
\begin{figure}
    \centering
    \begin{minipage}{0.3\linewidth}
        \centering
        \includegraphics[width=\linewidth]{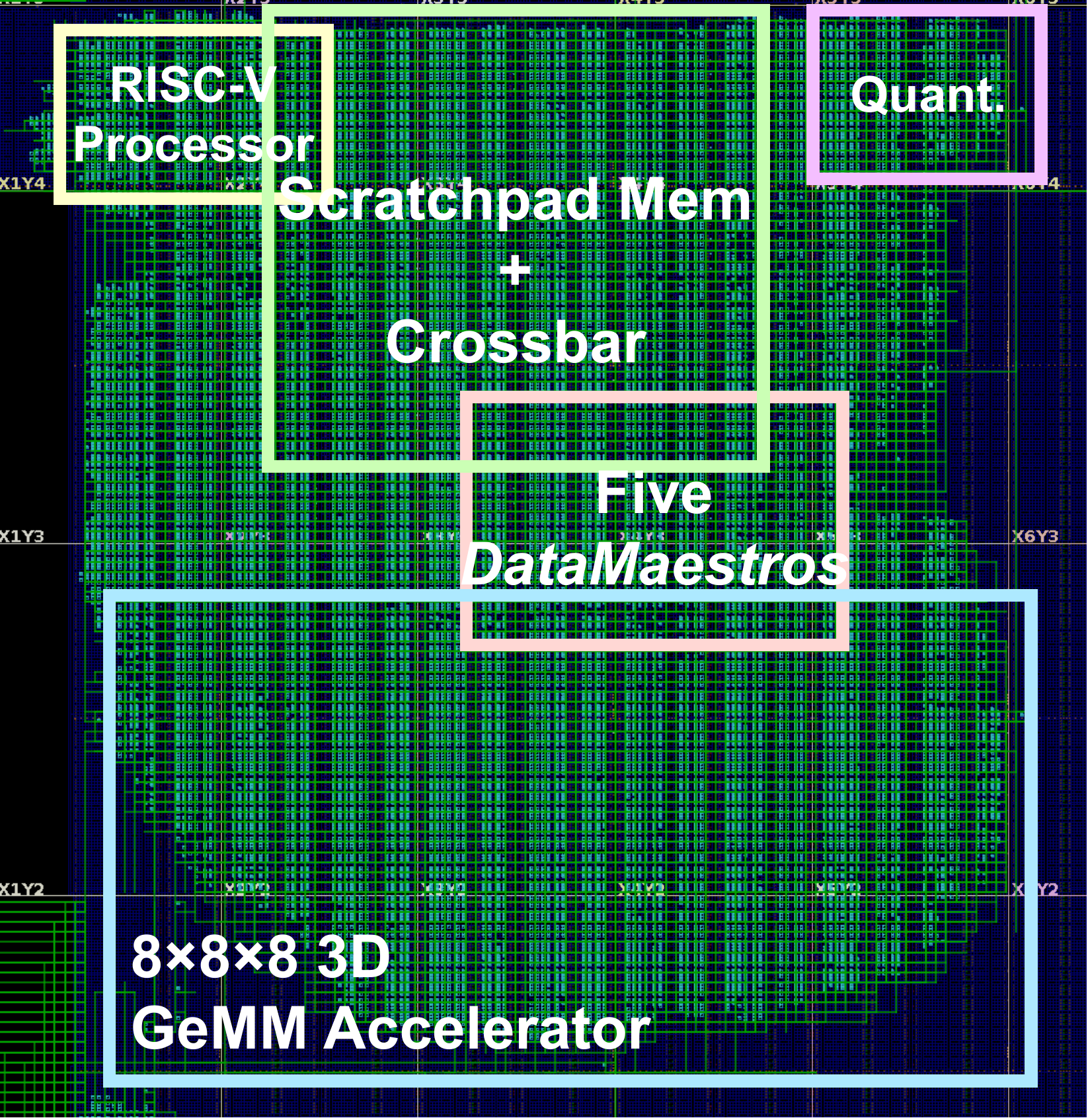}
    \end{minipage}
    \begin{minipage}{0.50\linewidth}
        \centering
        \begin{threeparttable}
            \footnotesize
            \begin{tabular}{c|c}
                \toprule
                \textbf{Platform}            & VPK180         \\
                \textbf{Clock Frequency}     & 125MHz         \\
                \hline
                \textbf{LUTs Total}          & 265k           \\
                \textbf{Regs Total}          & 59k            \\
                \textbf{LUTs GeMM}           & 124k (46.79\%) \\
                \textbf{Regs GeMM}           & 8k   (13.56\%) \\
                \textbf{LUTs \streamersname} & 14k  (5.28\%)  \\
                \textbf{Regs \streamersname} & 4.4k  (7.46\%) \\
                \bottomrule
            \end{tabular}
        \end{threeparttable}
    \end{minipage}
    \caption{The FPGA implementation (left) and resource utilization (right) of the \streamername evaluation system.}
    \label{fig: fpga_impl}
    \vspace{-0.4cm}
\end{figure}

We integrate our \streamername evaluation system into a customized SoC platform derived from \cite{paulin2024occamy} and prototype it on the AMD Versal\textsuperscript{\texttrademark} VPK180 FPGA, which facilitates rapid and convenient real-world DNN evaluations.
The annotated layout and resource utilization of our system are presented in Figure~\ref{fig: fpga_impl}.

\subsubsection{DNN Performance Benchmarking}
\label{sec:dnn_benchmark}
We benchmark ResNet-18 \cite{he2016identity}, VGG-16 \cite{simonyan2014very}, ViT-Base-16 \cite{dosovitskiy2020image}, and BERT-Base \cite{devlin2018bert} on our FPGA SoC system, with GeMM core utilization listed in Table \ref{tab:ai_workload}.
All four networks achieve utilization above $95\%$, with VGG-16 and ViT-Base-16 reaching nearly $100\%$. This high GeMM core utilization across these real-world DNN workloads highlights the efficacy of \streamername in minimizing data stream interruptions, thereby improving the performance of the accelerator system.

\begin{table}[h]
    \centering
    \begin{threeparttable}
        \caption{GeMM core utilization (in \%) of \streamername-boosted accelerator under real-world DNN workloads on VPK180 FPGA.}
        \label{tab:ai_workload}
        \begin{tabular}{lcccc}
            \toprule
                                              & \textbf{ResNet-18} & \textbf{VGG-16} & \textbf{ViT-B-16} & \textbf{BERT-Base} \\
            \midrule
            \textbf{Type}                     & CNN                & CNN             & Transformer       & Transformer        \\
            \textbf{Utilization (\%)}\tnote{$\dagger$} & 95.45              & 100.00          & 99.98             & 97.85              \\
            \bottomrule
        \end{tabular}
        \begin{tablenotes}
            \footnotesize
            \item \begin{minipage}[t]{\linewidth}[$\dagger$] Utilization is calculated as the ratio of theoretical computation cycles without memory stalls to the active cycles of \streamername.\end{minipage}
        \end{tablenotes}
    \end{threeparttable}
    \vspace{-0.5cm}
\end{table}

\subsection{Area and Power breakdown}
\label{sec: power_area_break}

\begin{figure}[t]
    \centering
    \includegraphics[width=0.95\linewidth]{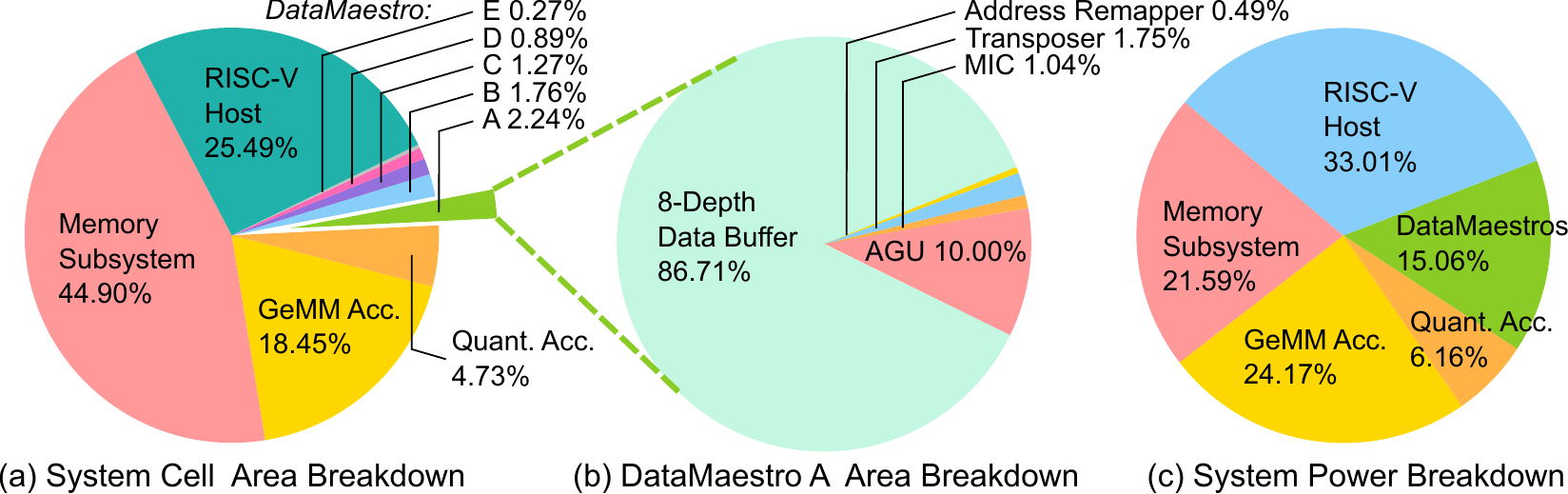}
    \caption{Breakdown of cell area (a) and total power (c) for the evaluation system, and area composition of \streamername $A$ (b).}
    \label{fig:area_power_breakdown}
    \vspace{-0.4cm}
\end{figure}

When synthesized using 22nm FDX technology, our system occupies $0.61 mm^{2}$ cell area and consumes $329.4 mW$ of total power when executing an $M$=$N$=$K$=$64$ GeMM (GeMM-64) workload at 1 GHz, achieving a total system-level energy efficiency of $2.57 \text{TOPS}/\text{W}$. The detailed system area and power breakdowns are shown in Figure \ref{fig:area_power_breakdown} (a) and (c), highlighting the five main components: the GeMM accelerator, Quantization accelerator, five \streamersname, the scratchpad memory subsystem, and the RISC-V host.
The first four components form the core of the accelerator system, accounting for $74.52\%$ of the total area and consuming $66.99\%$ of the total power.
The five \streamersname collectively occupy only $6.43\%$ of the system area and $15.06\%$ of the total power, with individual area usage ranging from $0.28\%$ to $2.33\%$.
The variation in area consumption demonstrates that \streamername's design-time parameters alter the deployed hardware to meet the specific requirements of each data stream, showcasing its high versatility.
Furthermore, to investigate the overhead of each feature inside \streamername, we conduct a detailed analysis of the area composition of a single \streamername $A$—the most advanced one among the five instantiations, as shown in Figure \ref{fig:area_power_breakdown} (b). The majority of the area is occupied by data FIFOs ($87.76\%$) and MICs ($1.04\%$) to support decoupled fine-grained prefetch.
The AGU, responsible for producing the $6$-$D$ temporal and $2$-$D$ spatial addresses, accounts for $10\%$ of the total area. The Transposer takes up $1.75\%$ of the total area. Lastly, the address remapper, which is essentially a multiplexer of permuted address bits, occupies a negligible $0.49\%$ of the area.

\subsection{State of the Art Comparison}

\subsubsection{Comparison with SotA DNN Dataflow Accelerators}
\begin{figure}[t]
    \centering
    \begin{minipage}{0.50\linewidth}
        \centering
        \includegraphics[width=\linewidth]{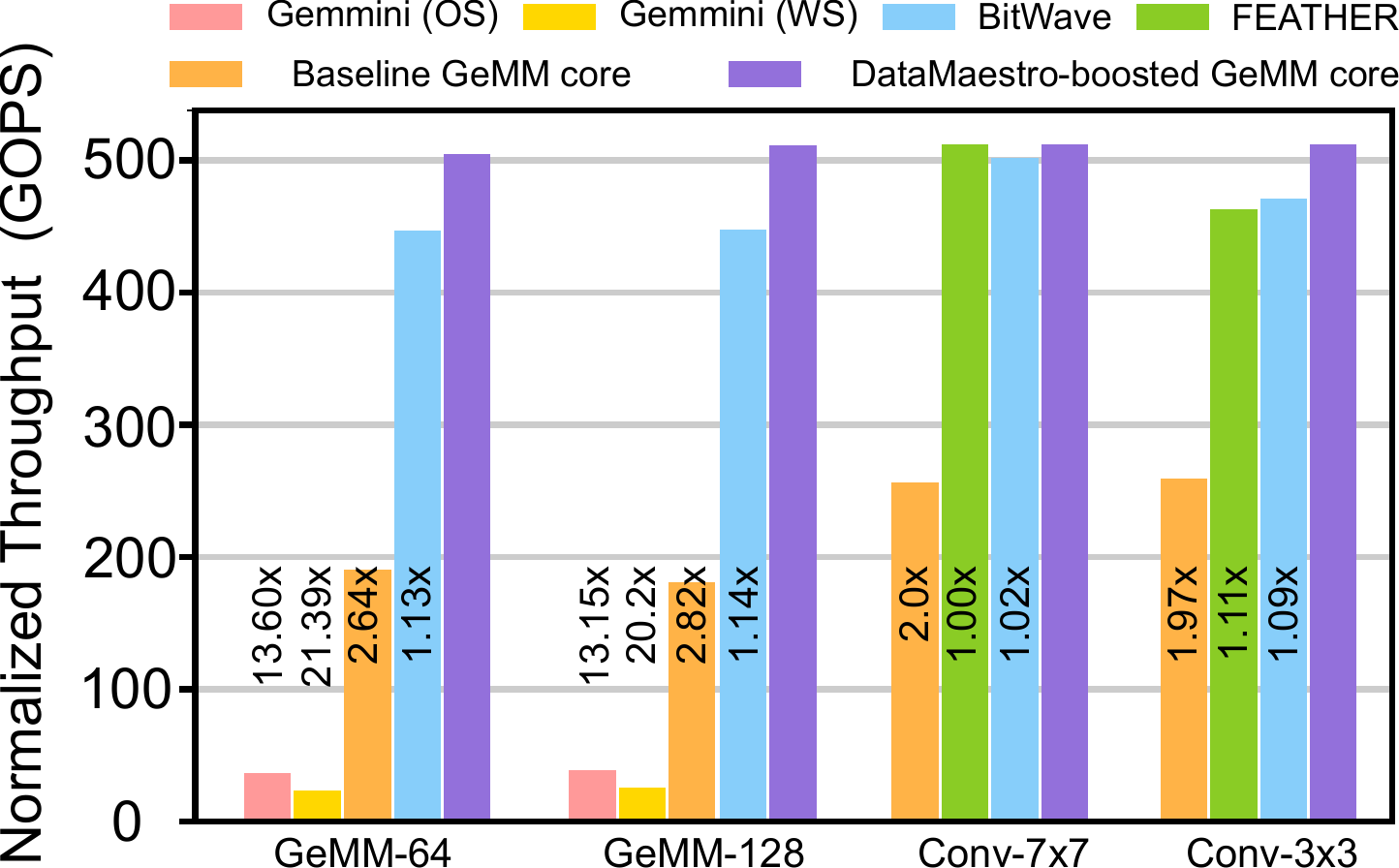}
    \end{minipage}
    \begin{minipage}{0.48\linewidth}
        \centering
        \scriptsize
        \begin{threeparttable}
            \begin{tabular}{c|c|c}
                \toprule
                                                    & Area              & Power              \\
                \hline
                Buffet \cite{pellauer2019buffets}   & $2\%$             & $14\%$             \\
                Softbrain \cite{nowatzki2017stream} & $4.3\%$           & $15.3\%$           \\
                Bitwave \cite{shi2024bitwave}       & $11.9\%$          & $25.5\%$           \\
                FEATHER \cite{tong2024feather}      & $8.9\%$           & N/A                \\
                \hline
                \textbf{\streamername}              & $\textbf{6.43\%}$ & $\textbf{15.06\%}$ \\
                \bottomrule
            \end{tabular}
        \end{threeparttable}
    \end{minipage}
    \caption{Comparison of normalized throughput (left) and data movement area and power cost ($\%$) inside the whole accelerator system (right) between SotA solutions and \streamername. Some data is not revealed in the literature; therefore, the corresponding bars are omitted in (a).}
    \label{fig: sota_comparison}
    \vspace{-0.5cm}
\end{figure}

We compare the throughput of our \streamername-boosted accelerator with SotA DNN dataflow accelerators, including Gemmini in output-stationary (OS) and weight-stationary (WS) modes \cite{genc2021gemmini, gonzalez202116mm}, FEATHER \cite{tong2024feather}, and BitWave \cite{shi2024bitwave} using representative DNN kernels.
For all systems, throughput is normalized to the same number of PEs (512) and clock frequency (1 GHz) of the accelerator for a fair comparison.
As the results in Figure \ref{fig: sota_comparison} (left) show, our \streamername-boosted accelerator achieves throughput gains ranging from $1.05\times$ to $21.39\times$ compared to SotA accelerators across diverse DNN kernels. This significant performance improvement results from the combined efforts of \streamername's design-time and runtime flexibility, fine-grained prefetch, and addressing mode switching techniques.

\subsubsection{Comparison with SotA Data Streaming Engines}
We further compare data streaming engines in Buffet\cite{pellauer2019buffets}, Softbrain\cite{nowatzki2017stream} and data movement units in BitWave\cite{shi2024bitwave}, and FEATHER\cite{tong2024feather} with \streamername and summarize their area and power overhead results in Figure \ref{fig: sota_comparison} (right). The five \streamersname in our evaluation system jointly occupy $5\%$ and $15.58\%$ of the system area and power consumption, which is competitive with SotA solutions, while flexibly and efficiently streamlining data to boost the dataflow accelerator's performance.

\section{Conclusion}

In this paper, we present \streamername, a versatile, efficient, and open-source data streaming unit that brings the decoupled access/execute architecture to DNN dataflow accelerators.
\streamername supports a programmable $N$-Dimensional access pattern to accommodate diverse workload types and dataflows, incorporates fine-grained prefetch and addressing mode switching to mitigate bank conflicts, and enables customizable on-the-fly data manipulation to reduce memory footprints and accesses. We integrate five \streamersname with a Tensor Core-like GeMM accelerator and a Quantization accelerator into a RISC-V host system. FPGA prototyping and VLSI synthesis results demonstrate that \streamername helps accelerators achieve nearly $100\%$ utilization, which is $1.05-21.39\times$ higher than SotA solutions, with area and energy cost of merely $6.43\%$ and $15.06\%$ of the total system.

\section*{Acknowledgment}

This project has been partly funded by the European Research Council (ERC) under grant agreement No. 101088865, the European Union's Horizon 2020 program (CONVOLVE) under grant agreement No. 101070374, the Flanders AI Research Program, and KU Leuven.

\clearpage
\bibliographystyle{IEEEtran}

{
	\small
	\bibliography{reference}
}

\end{document}